\documentclass[a4paper,twocolumn,
english,aps,pre,floatfix,showpacs]{revtex4}
\usepackage[T1]{fontenc}
\usepackage[latin1]{inputenc}
\usepackage{amsmath}
\usepackage{babel}
\usepackage{graphics}
\usepackage{amssymb}

\begin{document}
 
\title{Wavelet transforms in a critical interface 
model for Barkhausen noise
}

\author{S.L.A. \surname{de Queiroz}}

\email{sldq@if.ufrj.br}

\affiliation{Instituto de F\'\i sica, Universidade Federal do
Rio de Janeiro, Caixa Postal 68528, 21941-972
Rio de Janeiro RJ, Brazil}

\date{\today}
\begin{abstract}
We discuss the application of wavelet transforms to a critical  interface model, which is 
known to provide a good description of Barkhausen noise in soft ferromagnets.
The two-dimensional version of the model (one-dimensional interface) is considered,
mainly in the adiabatic limit of very slow driving. 
On length scales shorter than a crossover length (which grows with the strength of 
surface tension), the effective 
interface roughness exponent $\zeta$ is  $\simeq 1.20$, close to the expected value 
for the universality class of the quenched Edwards-Wilkinson model.    
We find that
the waiting times between avalanches are fully uncorrelated, as the wavelet 
transform of their autocorrelations scales as white noise. Similarly, detrended
size-size correlations give a white-noise wavelet transform. Consideration of
finite driving rates, still deep within the intermittent regime, shows
the wavelet transform of correlations scaling as $1/f^{1.5}$ for intermediate 
frequencies.
This behavior is ascribed to intra-avalanche correlations.   
\end{abstract}
\maketitle
\section{INTRODUCTION}
\label{intro}

In this paper we use wavelet concepts~\cite{daub92,{numrec,hurst98}} to discuss 
assorted properties of a single-interface model which has
been used in the description of Barkhausen  ``noise'' 
(BN)~\cite{umm95,{us,us2,bark3,rdist,trdist}}.
BN is an intermittent phenomenon which reflects the
dynamics of domain-wall motion in the central part of the hysteresis cycle
in soft ferromagnets (see Ref.~\onlinecite{dz05} for a review). 
By ramping an externally applied magnetic field, one causes
sudden turnings (avalanches) of groups of spins. The consequent changes in
magnetic flux induce a time-dependent electromotive force $V(t)$ on a coil
wrapped around the sample. Analysis of $V(t)$, assisted by suitable
theoretical modeling, provides insight into both the domain structure itself and 
its dynamical behavior. 
It has been proposed that BN is an illustration of ``self-organized
criticality''~\cite{bw90,cm91,obw94,umm95}, in the sense that a broad 
distribution of scales (i.e. avalanche sizes) is found within a wide range 
of variation of the external parameter, namely the applied magnetic field,
without any fine-tuning. The interface model studied here~\cite{umm95}
incorporates a self-regulating mechanism, in the form of a demagnetization 
factor. 

This way, real-space properties, e.g., interface roughness,
reflect the divergence of the system's natural length scale, as
it self-tunes its behavior to lie close to a second-order
(interface depinning) transition. In this context, the application of
wavelet transforms, which by construction incorporate multiple length
scales~\cite{daub92,numrec,hurst98}, is naturally suggested.

Also, when one considers the time series of intermittent events which
characterizes BN, a broad range of variation of $V(t)$ is shown, in correspondence
with the similarly wide distribution of avalanche sizes. 
Specifically considering the model of Ref.~\onlinecite{umm95}, it is 
known that the demagnetizing term is responsible for
the introduction of short-time negative (inter-avalanche) correlations 
(such correlations are observed in experiments as well)~\cite{umm95,trdist}. Thus,
a finite time scale  (``loading time'') is introduced, which coexists
alongside the broad distribution of $V(t)$.
The tool most frequently used in the analysis of BN time series is the Fourier power 
spectrum, 
i.e., the (cosine) Fourier transform of the time-time autocorrelation function of the 
signal $V(t)$~\cite{dz05,ks00,dz02}. BN power spectra exhibit distinct types of behavior
along different frequency ranges, reflecting the fact that finite ``internal'' times
play relevant roles. For instance, the loading times referred to above are expected
to influence the low-frequency end of the power spectrum, which pertains to 
inter-avalanche correlations, while the high-frequency tail relates to intra-avalanche 
ones. It has been stated that ``understanding the power spectrum of the magnetization 
noise is a long standing problem''~\cite{dz02}. 

Some existing applications of wavelet transforms to the analysis of 
$V(t)$~\cite{gp91,{mtww00,lm00}} mainly aim at demonstrating that the 
resulting spectra can
successfully distinguish between BN originating from physically distinct materials
(e.g., samples under differing amounts of internal stress).
Semi-empirical classification schemes have been proposed~\cite{{mtww00,lm00}}.
Wavelet (Haar) transforms~\cite{daub92} have also been employed in conjunction with 
standard Fourier
series, in order to produce higher-order power spectra of experimental data for 
$V(t)$~\cite{obw94,pwd98,pwd98b}. Analysis of the corresponding results
provides relevant evidence concerning correlations between events at different   
frequency scales. While in this work we shall deal only with first-order
transforms, in  Sec.~\ref{sec:pw} below we shall comment on possible connections
of our own findings to those of Refs.~\onlinecite{obw94,pwd98,pwd98b}.

The paper is organized as follows. In Sec.~\ref{sec:2} we recall pertinent aspects of the 
interface model used here, and of our calculational methods, as well as some basic 
features of wavelet transforms. In Sec.~\ref{sec:3} we consider the 
scaling of interface roughness configurations. In Sec.~\ref{sec:pw} we investigate 
properties extracted from time series, namely waiting-time and avalanche size 
correlations. Finally, in Sec.~\ref{conc}, concluding remarks are made.    

\section{Model and wavelet transforms} \label{sec:2}

\subsection{Single-interface model for BN}
\label{subsec:2a}
 
We use the single-interface model introduced in
Ref.~\onlinecite{umm95} for the description of BN. In line with
experimental procedure, the external field $H$
acting on the sample is assumed to increase linearly in time, therefore
its value is a measure of ``time''.  Initially, we consider the
adiabatic limit of a very slow driving rate, thus avalanches are
considered to be instantaneous (occurring at a fixed value of the external
field). In this simplified version, a plot of $V(t)$ against $t$ 
consists of a series of spikes of varying sizes, placed at non-uniform 
intervals. Generalizations for a finite driving rate may be 
devised~\cite{us2,tad99,wd03}; they are investigated in Subsection~\ref{sec:IVd}
below.

Simulations are performed on an $L_x \times L_y \times \infty$  geometry,
with the interface motion set along the infinite direction. Here we consider
$L_y=1$ (system  dimensionality $d=2$, interface dimensionality $d^\prime = 1$). Periodic 
boundary conditions are imposed at $x=0,\,L$~. 

The interface ($180$-degree domain wall separating spins parallel to the external field
from those antiparallel to it) is composed by $L$ discrete elements whose $x$ coordinates
are $x_i=i$, $i=1, \dots, L$, and whose (variable) heights above an arbitrary reference 
level are $h_i$. The simulation starts with a flat wall: $h_i=0$ for all $i$.
  
Each element $i$ of the interface experiences a force given by:
\begin{equation}
f_i=u(x_i,h_i)+{\kappa}\,\left[h_{i+1} + h_{i-1}- 2\,h_i\right]+H_e~,
\label{force}
\end{equation}
where
\begin{equation}
H_e =H -\eta M~.
\label{He}
\end{equation}
The first term on the right-hand side of Eq.~(\ref{force}) represents quenched disorder, 
and is drawn from a Gaussian distribution of zero mean and width $R$;
the intensity of surface tension is set by $\kappa$, and the effective field $H_e$
is the sum of a time-varying, spatially uniform, external field $H$ and a 
demagnetizing field which is taken to be proportional to
$M=(1/L)\sum^{L}_{i=1} h_i$, the magnetization  (per site) of the previously 
flipped spins for a lattice of transverse width $L$.
Here we mostly use $R=5.0$, $\kappa=1.0$, $\eta=0.005$, values for which fairly broad
distributions of avalanche sizes are obtained~\cite{rdist,us,us2,bark3}.
The exception is Sec.~\ref{sec:3}, where (for reasons to be explained),
we allow the surface tension  $\kappa$  to vary.

The dynamics goes as follows. For fixed $H$, starting from zero, the sites are 
examined sequentially; at those for which $f_i > 0$, $h_i$ is increased by one 
unit, with $M$ being  updated accordingly; the corresponding new value of $u$ is 
drawn. The whole  interface is swept as many times as necessary, until only sites 
with $f_i < 0$ are left, which marks the end of an avalanche. The external field is 
then increased until $f_i=0$ for at least one site. This is the threshold
of a new avalanche, which is triggered by the update of the site(s) with $f_i=0$, 
and so on.   

Because of the demagnetizing term,  the effective field $H_e$ at
first rises linearly with applied field $H$, and then, upon further increase
in $H$, saturates (apart from small fluctuations)  at a value rather close to the
critical external field for the corresponding model {\em without}
demagnetization~\cite{umm95,us}.

\subsection{Wavelets}
\label{subsec:2b}

Wavelets are characterized by a scale parameter, $a$, and a translation 
parameter, $b$, such that the wavelet basis, $\{\psi_{a;b}(x)\}$ can be entirely derived 
from a single function $\psi(x)$ through 
\begin{equation} 
\psi_{a;b}(x) = \psi\,\left(\frac{x-b}{a}\right)\quad .
\label{eq:wbasis}
\end{equation}
The wavelet transform of a function $f(x)$ is given by:
\begin{equation}
{\cal W}[f](a,b)=\frac{1}{\sqrt{a}}\int_{-\infty}^\infty 
\psi_{a;b}^\ast(x)\,f(x)\,dx\quad.  
\label{eq:wtransf}
\end{equation}
Here we shall use the Daubechies wavelet family~\cite{daub92,{numrec,hurst98}}.
These are real functions (appropriate in the present case where the input signal is 
always a real number, whether it be an interface height or a voltage); in the
{\em discrete} transform~\cite{numrec} implementation used here, the scales
$\{a\}$ are hierarchically distributed, i.e., $a_j=2^{-j}\,a_0$. We have
experimented with the Daubechies wavelets of orders~\cite{daub92} $4$, $12$, 
and $20$, and found that, similary to Ref.~\onlinecite{hurst98}, the quality of our 
results does not seem to depend on that. Therefore we have chosen the lowest order,
Daub4, for our calculations. 

It must be noted that the Daubechies wavelet filter coefficients used here incorporate 
{\em periodic} boundary conditions~\cite{numrec}. In the applications to be discussed, 
for each case we shall comment on the specific consequences of this constraint.

Furthermore, following Ref.~\onlinecite{hurst98}, we have chosen to average over
the translation parameters $b$, thereby arriving at a set of averaged wavelet 
coefficients to be denoted by $W[f](a)$. Among the several possible choices, we have 
found that averaging the squared coefficients tends to give smoother results than,
e.g., using absolute values~\cite{hurst98}. Thus, we define:
\begin{equation}
W[f](a)=\left[ \langle \left({\cal W}[f](a)\right)^2\rangle_b\right]^{1/2}\quad ,
\label{eq:awc}
\end{equation}
where $\langle \cdot\rangle_b$ stands for average over the translation parameters
$b$.   

\section{Real space properties: interface roughness}
\label{sec:3}
We begin by applying wavelet transforms to interface roughness data. 
The roughness $w_2$ of a fluctuating interface with $N$ elements is  
the position-averaged square  width of the interface height above an arbitrary 
reference level~\cite{adgr02,rkdvw03}: 
\begin{equation}
w_2  =N^{-1}\,\sum_{i=1}^N\left(h_i-\overline{h} \right)^2\ ,
\label{eq:rough1}
\end{equation}
where $\overline{h}$ is the average interface height. Self-affinity 
properties are 
expressed in the Hurst, or roughness exponent $\zeta$~\cite{bar95,svr95}:
\begin{equation}
\langle w_2 (L)\rangle \sim L^{2\zeta}\ \ ,
\label{eq:zeta}
\end{equation}
where angular brackets stand for averages over the ensemble of allowed interface 
configurations, and [for the $(1+1)$ dimensional systems which will be our main concern 
here] $L$ is the profile length. 

Numerical evidence has been given~\cite{rdist} that, as regards interface 
configuration aspects, the model  described here is in the quenched Edwards-Wilkinson 
universality class. 
Thus~\cite{les93,ma95,mbls98,rhk03} one expects $\zeta \simeq 1.25$  in $d=2$~. 
 
We have simulated BN through the evolution in time of the adiabatic, $d=2$ version of 
the model described above. 
Steady state, i.e., the stabilization of $H_e$ of Eq.~(\ref{He}) against
external field $H$, occurs after some $200$ events, for the range of 
parameters used here. In order to avoid start-up effects, here and in 
all subsequent sections we have 
used only steady-state data in our statistics.
At the end of each avalanche, we wavelet-transformed the 
instantaneous configuration of interface heights, i.e. the set of $\{h_i\},\ i=1, \dots 
, L$. As the avalanches progress, one gets a 
sampling of successive equilibrium configurations, which in turn provides us
with an ensemble of the corresponding wavelet coefficients. 
For each scale these are then translation-averaged, as explained above.

In this case, the periodic boundary conditions imposed at the interface extremities
are naturally consistent with those implicit in the wavelet transform, thus no
potential mismatch arises. 

For comparison with BN simulation data, we generated an artificial profile with 
$\zeta=1.25$, using  the random midpoint displacement algorithm~\cite{pjs92}. Although 
earlier applications of wavelet transforms to fractional Brownian motion  were restricted 
to $0 < \zeta <1$ in Ref.~\onlinecite{hurst98}, we found no technical impediments
in going above that upper limit.

\begin{figure}
{\centering \resizebox*{3.2in}{!}
{\includegraphics*{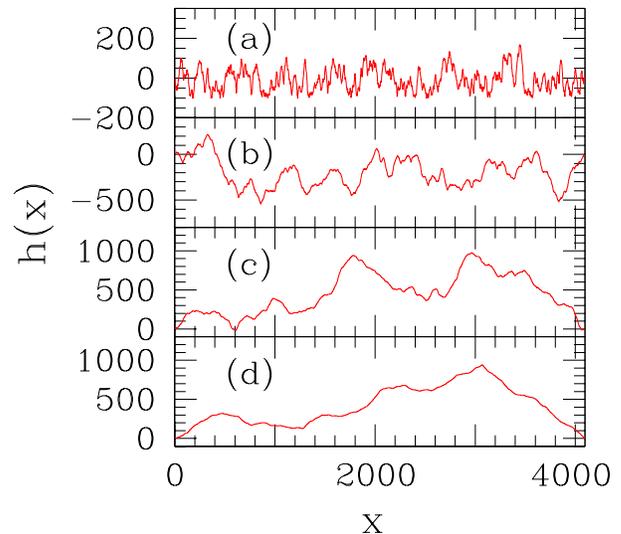}} \par}
\caption{(Color online) Snapshots of typical interface configurations. All with $4096$
sites, and periodic boundary conditions at the edges. (a)--(c): two-dimensional BN 
simulation, with varying surface tension (see Eq.~({\protect{\ref{force}}})), 
respectively: $\kappa=1.0$ (a), $3.0$ (b),
and $10.0$ (c). (d): artificial profile with $\zeta=1.25$.
}
\label{fig:prof}
\end{figure}

It is known that profiles with $\zeta>1$ are rather smooth~\cite{sw05}. This is
apparently at odds with the results to be expected from the force law, 
Eq.~(\ref{force}), from which the random locations of pinning centers would favor 
a rugged interface shape.
Thus, it is worth looking at interface configurations in real space. One anticipates
from Eq.~(\ref{force}) that the surface tension must play an important role
in this context. Accordingly, we allowed $\kappa$ to vary by one order of
magnitude. In Fig.~\ref{fig:prof}, one sees that on a fixed (system-wide) scale, the 
persistence trends
characteristic of $\zeta > 1/2$ are indeed reinforced by increasing $\kappa$.

\begin{figure}
{\centering \resizebox*{3.2in}{!}
{\includegraphics*{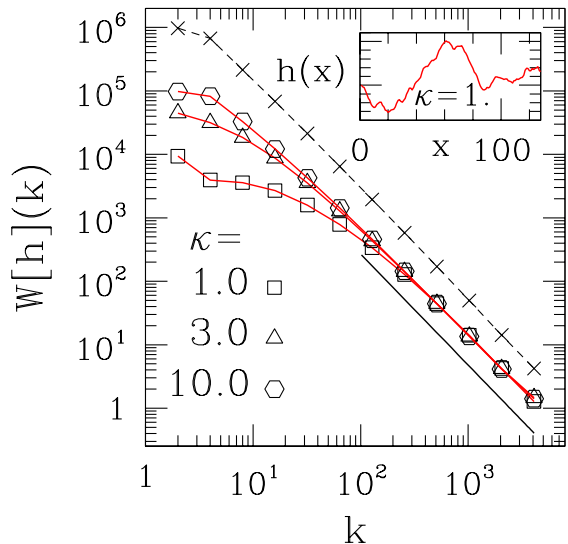}} \par}
\caption{(Color online) Double-logarithmic plot of averaged wavelet coefficients
against wavenumber $k$. Symbols joined by continuous lines: wavelet transform of 
interface roughness data from 
two-dimensional BN simulation (interface dimensionality $d^\prime=1$). $L=4096$, $10^5$ 
samples, with varying surface tension $\kappa$ (see 
Eq.~({\protect{\ref{force}}})). 
Crosses:  wavelet transform of 
synthetic profile with Hurst exponent $\zeta=1.25$. $L=4096$, $10^3$ samples.
Continuous line at bottom right has slope $-1.75$.
Inset: section of length $L^\prime=128$ of typical profile for $\kappa=1.0$, illustrating 
interface smoothness on short scales (compare Fig.~{\protect{\ref{fig:prof}}}~(a)).
}
\label{fig:r2d}
\end{figure}

One can have a quantitative understanding of the trends shown in Fig.~\ref{fig:prof},
with the help of wavelet transforms. The corresponding
results are displayed in Figure~\ref{fig:r2d}, where the horizontal axis
is in units of  inverse length scale, or~``wavenumber'' $k \equiv 1/a$. 
From scaling arguments~\cite{hurst98}, the averaged wavelet coefficients $W[h](k)$
for a self-similar profile are expected to vary as
\begin{equation}
W[h](k) \sim k^{-[(1/2)+\zeta]}\quad .
\label{eq:awc2}
\end{equation}
A least-squares  fit of a power-law dependence to the artificial-profile data for 
$64 \leq k \leq 4096$ gives $\zeta=1.25(1)$. Such a central estimate and its uncertainty 
are both in line with corresponding results for $0 <\zeta <1$~\cite{hurst98}. 

One sees that for BN data, $\zeta \simeq 1.25$ holds only up to a crossover scale,
which (as argued above) increases with $\kappa$. This is illustrated in the
inset of Fig.~\ref{fig:r2d}, where a section, with $1/32$ of the full length of the 
ragged  $\kappa=1.0$ interface of Fig.~\ref{fig:prof}~(a), is examined. On this scale, 
the profile is indeed much smoother than its parent.

A fit of $64\leq k \leq 4096$ data for $\kappa=10.0$ results in 
$\zeta=1.19(3)$. This can be  compared, e.g., with finite-size scaling estimates via 
Eq.~(\ref{eq:zeta}) for the present model with $\kappa=1.0$, and  a 
sequence of $400 \leq L \leq 1200$ with $O(10^6)$ 
configurations each, for which one quotes $\zeta=1.24(1)$~\cite{rdist}.   
Eq.~(\ref{eq:rough1}) reminds one that the latter method only considers fluctuations 
on short scales, thus in the present case it rightly captures the
persistent behavior
characteristic of that limit (at the expense of not being sensitive to the
different trends that dominate the picture at larger scales).  

We conclude that the quantitative behavior exhibited by interface roughness in BN is 
likely to change when studied on varying length scales. Though a regime 
should exist, which displays close similarity to the Edwards-Wilkinson class of 
interface evolution problems, this should cross over to a more ragged picture
on larger scales (the precise location of such change being determined by the
interplay between quenched randomness and surface tension). 
Wavelet transforms are thus a particularly suitable method for the study
of this problem, on account of the equal access that is provided to multiple length 
scales. 

\section{time series and power spectra}
\label{sec:pw}

\subsection{Introduction}
\label{sec:IVa}
As explained above, owing to the assumed linear increase of applied field  
with time (in analogy with experimental setups), we shall express time in
units of $H$ as given in Eqs.~(\ref{force}) and~(\ref{He}).

Initially we consider the adiabatic limit of very slow driving.

In experiment, the integrated signal $\int_{\Delta t} V(t)\,dt$ is
proportional to the magnetization change (number of upturned spins) during
the interval $\Delta t$. In the adiabatic approximation, a
box-like shape is implicitly assumed for each avalanche
(i.e. details of the internal structure of each peak, as it develops in time, are 
ignored, on account of its duration being very short),
thus the instantaneous signal intensity (spike height) is proportional to the
corresponding avalanche size.

As the signal is intermittent, there are significant periods (waiting
times, $WT$) of no activity at all. Waiting time distributions for the adiabatic regime 
were examined in Ref.~\cite{trdist}. These were found to 
be rather flat, apart from (i) a sharp cutoff at the high end (related to the finite
cutoff in the avalanche size probability distribution), and (ii) a number of
peaks concentrated in a somewhat narrow region, which are
associated to very frequent and small, spatially localized  (i.e., non-critical) events
involving typically $N = 1-10$ sites~\cite{us2}. 

\begin{figure}
{\centering \resizebox*{3.2in}{!}
{\includegraphics*{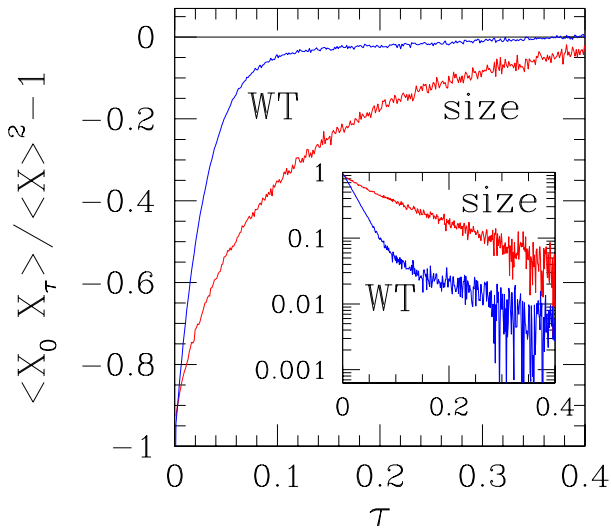}} \par}
\caption{(Color online) Waiting-time (WT) and size correlations (see 
Eq.~({\protect{\ref{eq:corr}}})) against ``time'' in the adiabatic regime,
for system with $L=400$, $2\times 10^4$ samples. Inset: absolute values of 
$G_X(\tau)$ on semi-logarithmic plot, same data range as in main Figure.   
}
\label{fig:wtscorr}
\end{figure}

We investigate the auto-correlations of two quantities, namely waiting times ($WT$), 
and avalanche sizes (i.e., BN spike voltages $V$). For $X=WT,\,V$
we calculate normalized, two-time connected correlations, averaged over $t$: 
\begin{equation}
G_X(\tau) \equiv \frac{\langle X(t)\,X(t+\tau) \rangle_t}{\langle X(t) \rangle^2_t} -1
\ .
\label{eq:corr}
\end{equation}
For a system with $L=400$, we have generated $2\times 10^4$ distinct time series of 
BN events. It is  known~\cite{umm95} that, on account of the demagnetizing factor, 
size-size correlations are negative at 
short times, and decay with a characteristic relaxation time 
which (for this system size, and for the values of physical parameters used here), is 
$\tau_0 \simeq 0.14$~\cite{trdist}. Thus, for each sample we calculated correlations in  
the range $0 \leq \tau \leq R$, $R=1.2$, by scanning moving ``windows'' of width $R$
along an interval of width $10 R$. In preparation for ulterior wavelet analysis,
the results were binned into $N=1024$ equal-width bins. Our results are depicted in 
Fig.~\ref{fig:wtscorr}.

The exponential behavior of  size data, noted earlier~\cite{umm95,trdist}, 
is clearly discernible in the Figure even for $\tau \gtrsim 0.3$, by which stage the 
signal-to-noise
ratio has dipped to something close to unity. Waiting-time 
correlations initially seem to follow a similar exponential trend (with a time constant 
$\simeq 1/4$ that for their size counterpart); however, a sharp ``shoulder''
develops at $\tau \approx 0.1$, signalling an abrupt end to the exponential regime.
This indicates that negative waiting-time and size correlations have
differing underlying causes. 

\subsection{Waiting-time correlations}
\label{sec:IVb}

Indeed, in calculating the correlations shown in Fig.~\ref{fig:wtscorr}, the time 
separation $\tau$ between any two waiting times is considered to be the
separation between their respective starting moments (the same is done for size 
correlations, but it turns out to be of no further consequence, as avalanches are 
instantaneous in the adiabatic regime).
This implies that the minimum separation between two waiting times is the extent of 
the shortest of the two. Therefore, an effect arises at very short times $\tau$, which 
is the analogue of hard-core repulsion for stoichiometric problems in real space.
Since the distribution of waiting times is flat on a logarithmic scale~\cite{trdist}
(thus $P(WT) \sim 1/WT$ on a linear scale), and assuming waiting times to be
uncorrelated (to be checked below), Eq.~(\ref{eq:corr}) gives $|G_{WT}(\tau)| \approx
1 -a\,\tau \simeq \exp(-a\tau)$ for $\tau \to 0$. 

In order to eliminate this artifact, 
we then decided to index waiting times simply by their order of occurrence, thus (with 
$j$, $k$ nonnegative integers)
\begin{equation}
G^{\,\prime}_{WT}(j)=\frac{\langle WT(k)\,WT(k+j)\rangle_k}{\langle WT(k)\rangle_k^2}-1\ 
.
\label{eq:corrwt}
\end{equation}
In analogy with our earlier procedure, correlations were accumulated for $j=1,\dots\,, 
N$ ($N=1024$) by generating $20$ independent series of $10N$ consecutive events; for each 
series  we scanned moving ``windows'', each comprising $N+1$ events, i.e., $N$ waiting 
times. This time, the result was essentially flat noise, with no apparent short-time 
structure (see inset in Figure~\ref{fig:twcorr} below). Therefore, further 
characterization must proceed via spectral analysis. We briefly recall how this
can be done using wavelets.  

Assume one has $1/f^\alpha$ noise. One calculates and
wavelet-transforms the corresponding ensemble-averaged autocorrelations, and then
translation-averages the resulting coefficients at each scale.  Denoting
the set of averaged wavelet coefficients by $\{ W[g](T)\}$, where
$\{T\}$ stands for the hierarchical set of wavelet timescales, and changing the 
independent variable to ``frequency'' $f=1/T$, one expects from 
scaling~\cite{hurst98,svr95}:
\begin{equation}
 W[g](f) \sim f^{-\alpha}\quad .
\label{eq:fscale}
\end{equation} 
For $\alpha >1$ this is derived immediately from Eq.~(\ref{eq:awc2}), plus
the exponent relation $\alpha=1+2\zeta$~\cite{rkdvw03,rdist}. Though for $0 \leq\alpha 
\leq 1$ the scaling of cumulants of the noise 
distribution differs from that for $\alpha>1$,
the basic scaling properties underlying Eq.~(\ref{eq:fscale}) remain valid~\cite{adgr02}.

Eq.~(\ref{eq:fscale}) can be tested with pure $1/f^\alpha$ noise
via the usual procedure of first producing a sequence of Gaussian
white noise, Fourier-transforming that sequence, multiplying the
Fourier components by $f^{-\alpha/2}$ and then inverting the
Fourier transform~\cite{adgr01,adgr02}. The resulting sequence is pure
$1/f^\alpha$ noise. An example with $\alpha=1/2$ is 
shown in Figure~\ref{fig:twcorr} below. 

\begin{figure}
{\centering \resizebox*{3.2in}{!}
{\includegraphics*{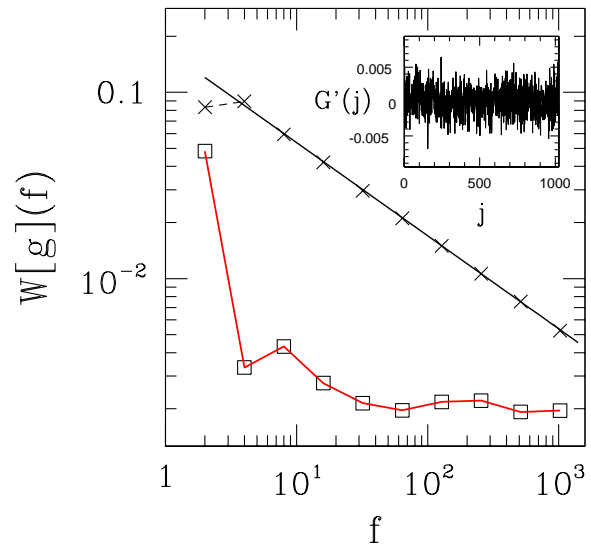}} \par}
\caption{(Color online) Double-logarithmic plot of averaged wavelet 
coefficients against frequency $f$. Squares: wavelet transform of waiting-time 
autocorrelation data from two-dimensional
BN simulation in the adiabatic regime, calculated according to 
Eq.~({\protect{\ref{eq:corrwt}}}). $L=400$, $20$ independent series
of $10 \times 1024$ waiting times. Crosses: wavelet transform of
autocorrelations for synthetic $1/f^\alpha$ noise, $\alpha=1/2$, $L=4096$, 
$5\times 10^3$ samples. A least-squares fit of
$16<f<1024$ data gives $\alpha=0.51(1)$. Continuous line has slope $-1/2$. Inset: 
waiting-time correlations from BN simulation, calculated according to 
Eq.~({\protect{\ref{eq:corrwt}}}).
}
\label{fig:twcorr}
\end{figure}

Our results for BN are shown in Fig.~\ref{fig:twcorr}. Apart from the lowest frequency 
scale (which is not expected to fall in line with the rest, as it represents 
the most smoothed-out behavior~\cite{numrec}), the flatness of 
the averaged coefficients against varying scales strongly indicates 
that $\alpha=0$ (white noise), i.e., waiting times are indeed uncorrelated.

The sequences of waiting-time correlation data of course need not be periodic. 
However, as seen above, they behave as random noise. Contrarily to, e.g., 
generalized brownian-motion profiles, such data are non-cumulative (i.e., 
they are not constrained in the fashion of consecutive positions of a random-walker, 
which cannot differ by more than one step length). Thus, the periodic boundary 
conditions implicit in the wavelet transform are not expected to introduce
significant distortions in their analysis. 

\subsection{Size correlations: adiabatic regime}
\label{sec:IVc}

We now turn to the treatment of voltage data. For the adiabatic version of the interface 
model, of course only inter-avalanche voltage correlations can be evaluated. 
As mentioned above, the data in Fig.~\ref{fig:wtscorr}
are very well fitted by an exponential, with a ``loading time'' $\tau_0=0.14(1)$. 
One then expects the Fourier power spectrum to be essentially flat for $f
\ll \tau_0^{-1}$, and 
to behave as $1/f^2$ for $f \gg \tau_0^{-1}$. This has indeed been found, e.g., in 
ref.~\cite{us2}.

The correlations to be wavelet-transformed are non-periodic and follow
a clear baseline trend,
therefore one needs to assess and eliminate potential distortions caused
by: (i) using a periodic wavelet basis, and (ii) the baseline trend
itself.

In Fourier analysis, the standard way to deal with (i) is by zero-padding
a region around the function to be transformed~\cite{numrec}.
However, zero-padding does not work well when
the function varies by orders of magnitude between the extremes
of the interval~\cite{numrec}, as is the case here where only
fluctuations are left at the upper end. Techniques have been
developed to remove the effects of periodic boundary conditions from
wavelet transforms (i.e., to consider ``wavelets on the 
interval'')~\cite{fp96}. These have very recently been 
translated into published computer code~\cite{numrec3},
restricted to the Daub4 class. In the following, motivated especially
by the need to address point (ii), we
propose a simplified approach based on detrending ideas.
Combinations of wavelet decomposition and detrending have been 
investigated~\cite{mpp05}; however, the averaged coefficient analysis 
which is our main concern here has not been considered, except for some
very simple cases (linear and quadratic drift~\cite{hurst98}). 

\begin{figure}
{\centering \resizebox*{3.2in}{!}
{\includegraphics*{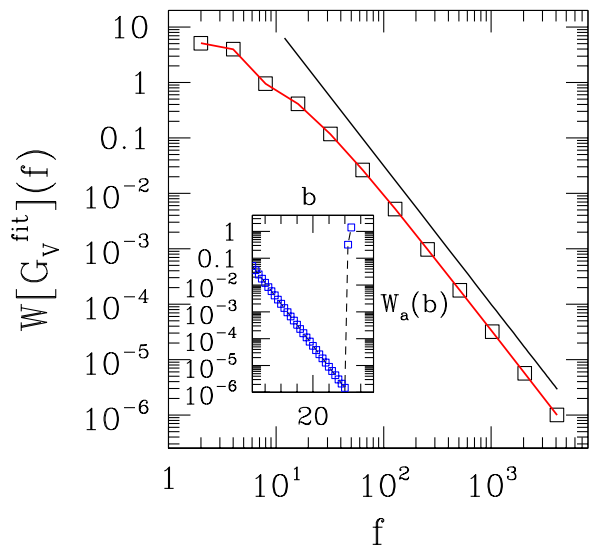}} \par}
\caption{(Color online) Main diagram: double-logarithmic plot of averaged wavelet 
coefficients against frequency $f$. Squares: wavelet transform of $N=4096$ points
of fitting function for size correlations, $G_V^{\,\rm fit}(\tau)$, for $0 < \tau < 1.5$. 
At each hierarchical 
level $j>2$, the last two coefficients were omitted from the averages (see text).
Continuous straight line has slope $-5/2$. Inset: semilogarithmic plot of
(absolute value of) all $32$ wavelet coefficients ${\cal W}[G_V^{\,\rm fit}](a,b)$ 
[denoted by $W_a(b)$] against translation parameter $b$, at hierarchical level $j=5$. 
}
\label{fig:adtcorr}
\end{figure}

We first illustrate how the averaged coefficients are affected by an
overall exponential trend. Using the periodic Daub4 basis, we
wavelet-transformed the
size-correlation fitting function, $G_V^{\,\rm
fit}(\tau)=-\exp(-\tau/\tau_0)$. 
From Eq.~(\ref{eq:wtransf}), one has: 
\begin{equation}
{\cal W}[G_V^{\,\rm fit}](a,b)=\frac{1}{\sqrt{a}}\int_{-\infty}^\infty 
\psi_{a;b}(x)\,e^{-x/\tau_0}\,dx\ .  
\label{eq:wtexp}
\end{equation}
By changing variables, Eq.~(\ref{eq:wtexp}) turns into:
\begin{equation}
{\cal W}[G_V^{\,\rm fit}](a,b)=\sqrt{a}\,e^{-b/\tau_0} \int_{-\infty}^\infty
\psi_{1;0}(x^\prime)\,e^{-ax^\prime/\tau_0}\,dx^\prime\ .
\label{eq:wtexp2}
\end{equation}
The first $p=M/2$ moments (starting at zeroth order) of
Daubechies wavelets  of order $M$ vanish~\cite{numrec}. Thus, for 
$M=4$ as is the case here, Taylor-expanding the exponential 
in the integrand of Eq.~(\ref{eq:wtexp2}), one sees that the lowest-order
non-zero term is proportional to $a^{5/2}$, i.e.,
\begin{equation}
{\cal W}[G_V^{\,\rm fit}](a,b) \propto {a}^{5/2}\,e^{-b/\tau_0} + {\cal 
O} ({a}^{7/2})\ \ .
\label{eq:wtexp3}
\end{equation}
We evaluated $G_V^{\,\rm fit}(\tau)$ at $N=4096$ equally-spaced points in the interval 
$0 < \tau < 1.5$, and wavelet-transformed it. 
For each hierarchical level $j>2$, we plotted all $2^j$ wavelet coefficients, and found 
that they fall on the exponential-decay pattern of the original function, and
(at the $j$--th hierarchical level) are proportional to $2^{-5j/2}$, both features as 
predicted in Eq.~(\ref{eq:wtexp3}), 
except for the last two (``wraparound'' coefficients~\cite{numrec}). In order to fulfill 
the implicitly assumed periodicity of the 
original function, the latter coefficients take values $\sim 10^j$ larger than the last 
preceding one (see an example for $j=5$ in the inset of Fig.~\ref{fig:adtcorr}). 
Including these data in the coefficient-averaging procedure
would introduce sizeable distortions (we did it, and found that the 
coefficients thus averaged behave as $1/f$, which is in clear disagreement with
the prediction of Eq.~(\ref{eq:wtexp3}) of a scaling power $5/2$).

To correct this artifact, we discarded the wraparound coefficients from
the averaging procedure. Similar
procedures have been adopted elsewhere~\cite{mpp05}.  
As can be seen in Fig.~\ref{fig:adtcorr}, this was enough to restore the expected 
behavior. Thus, point (i) above has been dealt with.
We also wavelet-transformed $G_V^{\,\rm fit}(\tau)$ using 
the periodic Daub12 basis. As expected, the coeficients behaved 
approximately as ${a}^{13/2}\,e^{-b/\tau_0}$. 
The last four coefficients
at each hierarchical level showed considerable increase against the exponential-decay
pattern (as opposed to the last two for Daub4). In summary, as regards
point (ii) we have shown
that the most prominent feature of the wavelet transform
(in the context of average wavelet coefficient scaling), namely the
Hurst-like exponent, of such a smooth function as the exponential fit is   
in fact basis-dependent.

\begin{figure}
{\centering \resizebox*{3.2in}{!}
{\includegraphics*{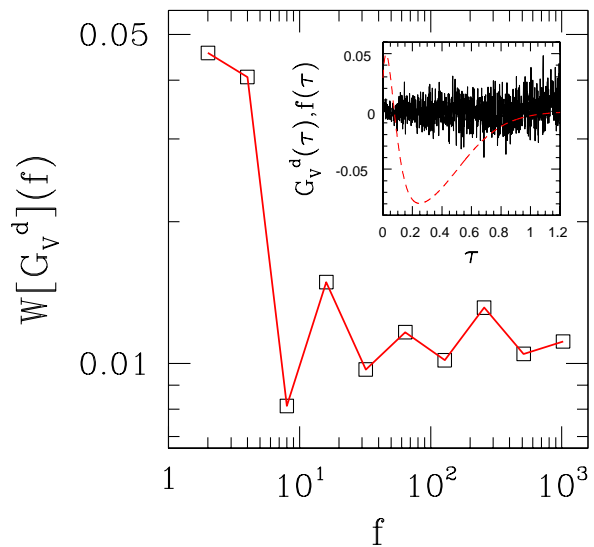}} \par}
\caption{(Color online) Double-logarithmic plot of averaged wavelet 
coefficients against frequency $f$. Squares: wavelet transform of 
detrended size autocorrelation data, $G_V^{\,\rm d}(\tau)$ from two-dimensional
BN simulation in the adiabatic regime. $L=400$, $2\times 10^4$ samples.
Inset: full lines: fully-detrended size correlations from BN simulation; 
dashed line: secondary adjusting functon $f(\tau)$ (see text).
}
\label{fig:dtscorr}
\end{figure}

Thus, our simulational data must be detrended in order to eliminate 
distortions coming from the smooth baseline, which risk contaminating all scales. 
We did this by first subtracting the dominant
exponential behavior given by $G_V^{\,\rm fit}(\tau)$; for further refinement, we 
then removed some remaining non-monotonic mismatch via least-squares fit of a secondary 
adjusting function $f(\tau)$ (a fourth-degree  polynomial enveloped by a single 
exponential),  so $G_V^{\,\rm d}(\tau)=
G_V(\tau)-G_V^{\,\rm fit}(\tau)-f(\tau)$. The result of wavelet-transforming the
fully detrended correlations is depicted in Fig.~\ref{fig:dtscorr}, while the 
corresponding  raw (detrended) data are shown in the inset of the same Figure (together 
with $f(\tau)$, so one
can have a quantitative estimate of how far the single-exponential fit goes to
describe the un-detrended data).
Note that $f(\tau)$ has significant smooth variations on scales of
$\delta\tau=0.05$ or longer, which translate into wavevectors $k \lesssim 
32$. We have wavelet-transformed partially-detrended data [$\,$i.e., 
without subtracting $f(\tau)\,$]. The respective averaged wavelet
coefficients are $\sim 10$ times larger than those for the fully-detrended
curve for $k \leq 16$, and fall fast for increasing $k$: at $k=64$ the 
ratio is $1.4$, and for $k>64$ both sets coincide to within less than 
$1\%$.
So, failing to subtract $f(\tau)$ introduces artificially large
coefficients at large scales, which are not noise-related.

Note that similar remarks apply
here as in the earlier case of waiting-time correlations, namely, since $G_V^{\,\rm 
d}(\tau)$ is essentially noise around a horizontal baseline, the periodic boundary
conditions implicit in the wavelet transform must not imply any significant distortion
in our results.  

The results exhibited in the main diagram of Fig.~\ref{fig:dtscorr} strongly indicate 
that the detrended size correlations behave as $1/f^0$ (white) noise. We defer discussion
of this until the next Subsection, where departures from the 
adiabatic regime are investigated. 

\subsection{Size correlations: finite driving rate}
\label{sec:IVd}

In order to discuss intra-avalanche correlations, one must introduce a finite
driving rate~\cite{us2,tad99,wd03}, so separate events within the same
avalanche can be ascribed to different instants in time. In line with standard 
practice~\cite{tad99,pds99,bt00,btn00,ks00} our basic
time unit is one lattice sweep, during which the external field is kept
constant, and all spins on the interface are probed sequentially as
described above.  In the adiabatic regime, the external field
is kept constant for the whole duration of an avalanche, i.e. for as many
sweeps as it takes until
no unstable sites are found along the interface. At finite driving rates, the
field is increased by a fixed amount, henceforth denoted $\Delta$, at the
start of each sweep
while an avalanche is taking place. Eventually, no more unstable sites
will be left, and then one proceeds as in the adiabatic regime, increasing
the field by the minimum amount $\delta H$ necessary to start a new avalanche.
In these ``real'' time units, the waiting time between the end of one event
and the start of the next is then $\delta H /\Delta$; however, in order to produce 
meaningful comparisons, especially between data acquired in the adiabatic and 
non-adiabatic regimes, it will be useful to keep referring to the ``absolute'' scale
given by the applied field $H$ itself, which unequivocally locates events along the
hysteresis cycle. 

As $\Delta$ grows,
the intermittent character of events is gradually lost as more and more avalanches 
coalesce~\cite{us2}, and one eventually crosses over to a regime in which the
interface is fully depinned, i.e. it moves at non-zero average speed.

\begin{figure}
{\centering \resizebox*{3.2in}{!}
{\includegraphics*{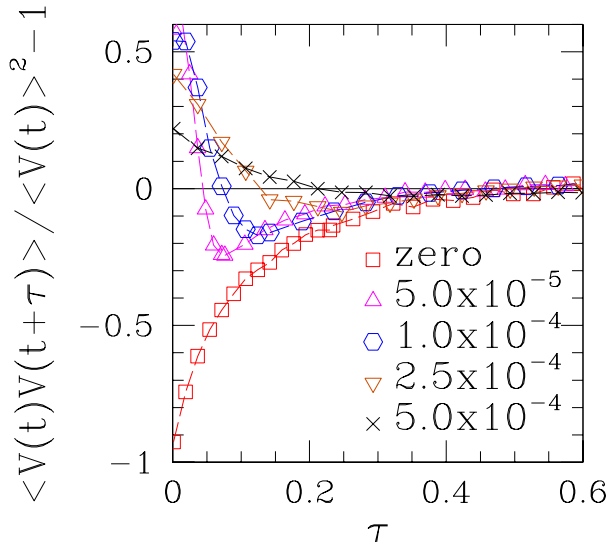}} \par}
\caption{(Color online) Normalized two-time correlations  (averaged 
over $t$) $\langle V(t)\,V(t+\tau) \rangle/\langle V(t) \rangle^2 -1$
from two-dimensional BN simulation, for system with $L=400$, and
driving rates $\Delta$ as given in key to symbols ($\Delta=0$
corresponds to adiabatic limit). ``Time'' is given in applied field units, i.e.
``absolute'' scale (see text).  
}
\label{fig:tcorr1}
\end{figure}

In Fig.~\ref{fig:tcorr1} we show autocorrelations for driving rates still within
the intermittent regime, compared with those for the adiabatic limit. The most 
significant change upon increasing $\Delta$ is the effective loss of negative short-time
correlations. In fact, this represents an excess of positive intra-avalanche
contributions, on top of the negative inter-avalanche terms (and some intra-avalanche 
ones as well) which still exist for non-zero $\Delta$ (on account of the demagnetizing
factor). Positive reinforcements arise mostly because, when many sites are overturned
during one lattice sweep, that same number of new sites will be probed by the interface.
For each new site, the quenched randomness term in Eq.~\ref{force} may, or may not,
contribute to further motion with roughly equal chances. By contrast, at a site which
remains pinned during one sweep, the interface stands fewer chances of getting unstuck,
as the contribution from the randomness term is kept constant;
depinning of such a site is more likely to happen if the field is substantially 
increased, i.e., during a subsequent avalanche. 

\begin{figure}
{\centering \resizebox*{3.2in}{!}
{\includegraphics*{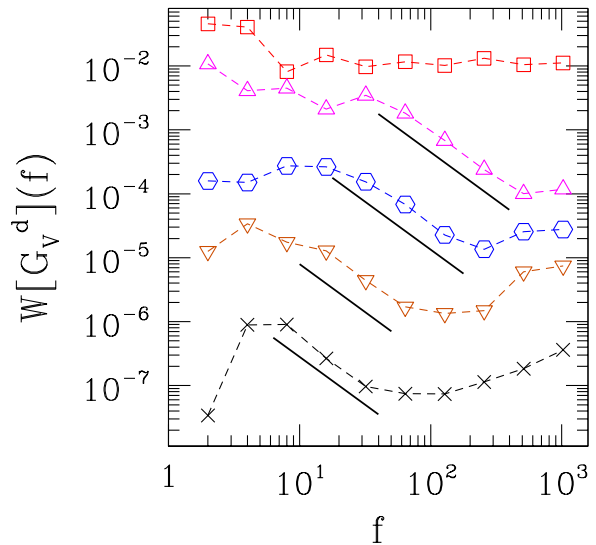}} \par}
\caption{(Color online)  Double-logarithmic plot of averaged wavelet 
coefficients against frequency $f$, from wavelet transform of 
detrended size autocorrelation data, $G_V^{\,\rm d}(\tau)$, for two-dimensional
BN simulations of system with $L=400$, and
assorted driving rates $\Delta$. Key to symbols is the same as in 
Fig.~{\protect{\ref{fig:tcorr1}}} ($\Delta=0$
corresponds to adiabatic limit). Frequency is given in inverse applied field units, i.e.
``absolute'' scale (see text).  
 Plots successively shifted downward by a factor of $10$ 
on vertical scale, to avoid superposition. Straight line segments 
mark subsets of $\Delta \neq 0$ regime where approximate $1/f^{1.5}$ behavior holds.
}
\label{fig:dtscorr2}
\end{figure}

We detrended the $\Delta \neq 0$ data of Fig.~\ref{fig:tcorr1} by similar procedures
to those used earlier for $\Delta=0$. The main difference was that detrending was done
in a single stage, fitting  $f(\tau)$ described in 
Subsec.~\ref{sec:IVc} to the raw data, and then subtracting the least-squares-fit from 
the original data. The results of wavelet-transforming the detrended data 
are shown in Figure~\ref{fig:dtscorr2}.

\begin{figure}
{\centering \resizebox*{3.2in}{!}
{\includegraphics*{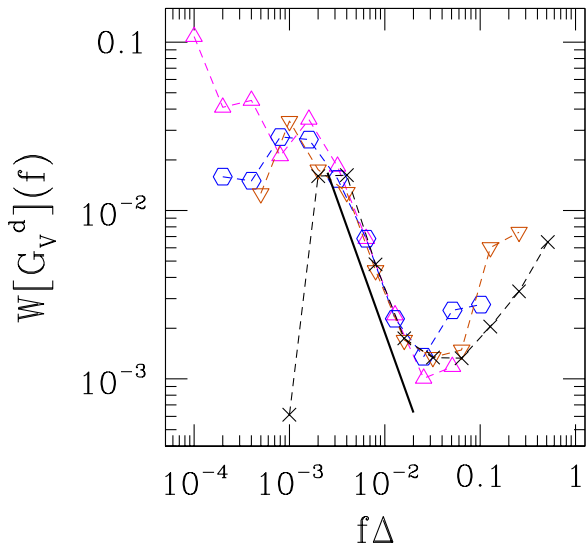}} \par}
\caption{(Color online)  Double-logarithmic scaling plot of averaged wavelet 
coefficients against ``natural'' frequency $f^\prime \equiv f\,\Delta$, from wavelet 
transform of 
detrended size autocorrelation data, $G_V^{\,\rm d}(\tau)$, for two-dimensional
BN simulations of system with $L=400$, and
assorted driving rates $\Delta \neq 0$. Key to symbols is the same as in
Figs.~{\protect{\ref{fig:tcorr1}}} and~{\protect{\ref{fig:dtscorr2}}}.
Full straight line has slope $-1.5$~. 
}
\label{fig:dtscorr3}
\end{figure}

One can see that, as opposed to the adiabatic regime, data for finite driving rates 
clearly exhibit a downward trend for a range of intermediate frequencies, spanning 
3-4 hierarchical levels, and which is characterized by an approximate $1/f^{1.5}$
behavior (the straight line segments in the Figure have slope $-1.5$).
Furthermore, with the ``absolute'' frequency $f$ given in 
inverse applied field units, and $\Delta$ given in units of applied field change
per unit time, dimensional arguments show that $f^\prime \equiv f\,\Delta$ is
the ``natural'' frequency variable (i.e., inverse ``real'' time). This is shown more 
clearly on a scaling plot, Figure~\ref{fig:dtscorr3}, where use of $f^\prime$ as the 
independent variable causes the $1/f^{1.5}$ sections of all $\Delta \neq 0$ data
to collapse.

Given that, in these slow- (but non-adiabatic) driving regimes, avalanche coalescence 
comprises only a small fraction of events~\cite{us2}, one can say that approximately
the same sequence of avalanches occurs for all $\Delta$ investigated here, only at
different ``real'' paces. Since the ``real'' time interval between
consecutive avalanches is $\delta h/\Delta$, and assuming $\delta h$ to be the same,
for different values of $\Delta$,
between two given avalanches (for the reasons just mentioned), one sees
that inter-avalanche correlations will shift to higher ``real'' frequencies as $\Delta$
grows. On the other hand, within a given avalanche, two sub-events separated by a given 
number of lattice sweeps are (by definition used in the simulation) separated by the same 
``real'' time  interval, thus their correlations are not shifted in ``real'' frequency 
for varying $\Delta$. Therefore we conclude that the collapsing sections of the
scaling plot correspond mainly to intra-avalanche correlations. 

First-order (Haar) spectra of  experimental BN data show that, for
Fe$_{21}$Co$_{64}$B$_{15}$  the high-frequency section falls initially as $f^{-1.2}$, 
and then crosses over to $f^{-1.9}$, while for Fe$\,$Si the decay is with  
$f^{-1.65}$~\cite{pwd98,pwd98b}. Though the exponent values in both cases are not
too dissimilar to the one found here, analysis of higher-order spectra~\cite{obw94}
leads to a more nuanced picture. For Fe$_{21}$Co$_{64}$B$_{15}$, it is found 
that most of the power in the high-frequency range comes from intra-pulse 
correlations~\cite{pwd98}, 
similar to our conclusion above, whereas for Fe$\,$Si the conclusion was that the 
high-frequency power is mainly connected to the inter-pulse sort~\cite{pwd98b}.   
Therefore it would appear that the dynamics of the present 
model is closer to that of BN in materials like Fe$_{21}$Co$_{64}$B$_{15}$ than in
Fe$\,$Si. 

\section{Discussion and Conclusions}
\label{conc}
We have discussed the application of wavelet transforms to the description of
both real-space and time-like properties of an interface model, which is used for the
description of Barkhausen noise in soft ferromagnets. Most of our calculations
involved the scaling properties of positional averages of wavelet coefficients, taken at 
each  hierarchical (size) level, as first proposed in Ref.~\onlinecite{hurst98}.
In some instances we showed that direct analysis of individual
coefficients was
called for, in order to unravel artificial effects which would otherwise distort
our aggregate results.

Here we considered the $d=2$ version  of the model (thus the
interface dimensionality is $d^\prime=1$), mainly in the adiabatic limit of very
slow driving, for which the sudden ``avalanches'' of domain wall motions are
considered to occur instantaneously. In Subsection~\ref{sec:IVd}, we extended
our study to finite driving rates, in order to analyze intra-avalanche correlations 

Our investigation of real-space aspects consisted in the evaluation of
the characteristic interface roughness exponent $\zeta$. On scales shorter
than a crossover length (which turns larger as the intensity of surface tension
grows), we get $\zeta=1.20(3)$, close to
$\zeta=1.24(1)$, derived by other methods for the same 
model~\cite{rdist}, and also to assorted estimates
for quenched Edwards-Wilkinson systems~\cite{les93,ma95,mbls98,rhk03}, 
which give $\zeta \simeq 1.25$. 

Turning to time series, in Subsection~\ref{sec:IVb} we showed that a proper
indexation of the sequence of waiting-times between avalanches is crucial,
in order to avoid artificial short-time negative correlations. Procedures
similar to that  used here,
namely, indexing waiting times simply by their order of occurrence (instead of using
the starting time of each interval), have been used consistently in the
context of self-organized criticality scaling~\cite{dp02}. Our final result
(see Fig.~\ref{fig:twcorr}) was that the correlations between waiting times are white 
noise, i.e., these quantities are fully uncorrelated. Going back to the
rules of interface motion, and to Eqs.~(\ref{force}) and~(\ref{He}), one sees that
this is a signature of the quenched-randomness term $u(x_i,h_i)$. This 
fact is in contrast to the behavior of size correlations, which are strongly influenced 
by demagnetization~\cite{umm95,trdist}.

In Subsection~\ref{sec:IVc}, we started from the known fact that, in the adiabatic 
regime, size-size correlations are negative at short times, and decay approximately
as an exponential~\cite{umm95,trdist}. 
By direct analysis of (non-averaged) wavelet coefficients, we illustrated 
practical ways to deal with artifacts introduced by the periodicity of the
wavelet basis used. It turned out
that the smooth baseline function, to which noise data are fitted, can introduce
distortions at all levels of the wavelet transform. Furthermore, such distortions
are non-universal in the sense that they depend on the wavelet basis.
Thus, in order to obtain meaningful results from  averaged wavelet
coefficients, one must fully detrend the raw data. 
Once we did so, we found strong
indications that the detrended size correlations behave as white noise (see
Fig.~\ref{fig:dtscorr}). This is apparently at odds with earlier (Fourier)
power-spectrum results (see, e.g., Ref.~\onlinecite{us2} and references therein), which 
would lead one to expect $1/f^2$ behavior, at least for high frequencies.
However, the derivation of the latter result (e.g., by direct integration) 
fully takes into account the exponential baseline shape, thus one is referring
to a different object. Here, as explained above, we are dealing with detrended
data.

Finally, in Subsection~\ref{sec:IVd}, we considered size correlations against time in 
non-adiabatic regimes (but well within the driving-rate range where intermittency 
still holds~\cite{us2}). For driving rates $\Delta$ spanning one order of magnitude,
we  found rather well-defined frequency intervals for which
detrended correlations behave as $f^{-\alpha}$, $\alpha \approx 1.5$, 
By changing variables from ``absolute'' to ``natural'' frequency, we found
that said intervals collapse together, which indicates
that they pertain to intra-avalanche correlations.
Rather than attaching much significance to the numerical value of the
power-law exponent (since the shortness of the interval along which such
behavior holds prevents one from doing so), one must emphasize the good
degree of curve collapse exactly in that section, and only there. This
indicates that  this section is the "special" one, i.e. it corresponds to
the frequency range along which universal (driving-rate independent)
properties hold. Furthermore, our considerations leading to the conclusion
that such scaling behavior reflects intra-avalanche correlations are
completely independent of  
the analysis of higher-order power spectra experimental data, carried out
in Refs.~\onlinecite{pwd98,pwd98b}, and which leads to the very same
conclusion as regards  BN in samples of Fe$_{21}$Co$_{64}$B$_{15}$. 
\begin{acknowledgments}

This research  was  partially supported by
the Brazilian agencies CNPq (Grant No. 30.6302/2006-3), FAPERJ (Grant
No. E26--152.195/2002), and Instituto do Mil\^enio de
Nanoci\^encias--CNPq.
\end{acknowledgments}

\bibliography{biblio}  
\end{document}